\documentclass[conference,a4paper]{APSIPA2025}
\usepackage{amsmath}
\usepackage{booktabs} 
\usepackage{graphicx}
\usepackage{pgfplots}
\pgfplotsset{compat=1.18}
\usepackage{pgfplotstable}
\usepackage{amssymb}
\let\labelindent\relax
\usepackage{enumitem}
\usepackage[backend=biber,style=ieee,maxnames=99]{biblatex}
\usepackage{geometry}
\usepackage{fancyhdr}

\fancypagestyle{firststyle}{
  \fancyhf{}
  \fancyhead[C]{2025 Asia Pacific Signal and Information Processing Association Annual Summit and Conference (APSIPA ASC)}
}

\begin{document}

\title{Accelerated Convolutive Transfer Function-Based Multichannel NMF Using Iterative Source Steering}

\author{
\authorblockN{
Xuemai Xie\authorrefmark{1}, 
 Xianrui Wang\authorrefmark{1},
Liyuan Zhang\authorrefmark{1},
Yichen Yang\authorrefmark{1} and
Shoji Makino\authorrefmark{1}
}

\authorblockA{
\authorrefmark{1}
Waseda University, Japan \\
E-mails: xuemaixie@ruri.waseda.jp, wangxianrui97@gmail.com \\
ly.zhang@akane.waseda.jp, yang\_yichen@mail.nwpu.edu.cn \\
s.makino@waseda.jp}


}

\maketitle
\thispagestyle{firststyle}
\pagestyle{fancy}

\begin{abstract}
Among numerous blind source separation (BSS) methods, convolutive transfer function-based multichannel non-negative matrix factorization (CTF-MNMF) has demonstrated strong performance in highly reverberant environments by modeling multi-frame correlations of delayed source signals. However, its practical deployment is hindered by the high computational cost associated with the iterative projection (IP) update rule, which requires matrix inversion for each source. To address this issue, we propose an efficient variant of CTF-MNMF that integrates iterative source steering (ISS), a matrix inversion-free update rule for separation filters. Experimental results show that the proposed method achieves comparable or superior separation performance to the original CTF-MNMF, while significantly reducing the computational complexity.
\end{abstract}
\begin{IEEEkeywords}
Blind source separation, nonnegative matrix factorization, convolutive transfer function, fast algorithm.
\end{IEEEkeywords}

\section{Introduction}
Blind source separation (BSS) is a technique that recovers the source signals from only the observed sensor signals, without prior knowledge of the mixing process or the source characteristics \cite{makino2018audio, makino2007blind, wang2023multiple}. Based on how signals are mixed, BSS can be categorized into two types: instantaneous BSS and convolutional BSS~\cite{douglas2004natural, buchner2004generalization, yang2022iva}. Although the instantaneous mixing model is computationally efficient and conceptually simple, it fails to adequately capture real-world reverberation structures, resulting in terrible performance in environments with pronounced delays and reflections~\cite{makino2018audio, hyvarinen2001independent,wang2023spatially}.

As popular instantaneous BSS techniques, auxiliary function-based independent vector analysis (AuxIVA) \cite{kim2006independent, ono2011stable, hiroe2006solution,yang2023eusipco} and independent low-rank matrix analysis (ILRMA) \cite{kitamura2016determined} are widely used due to their stable separation performance. ILRMA replaces the source model used in AuxIVA with the nonnegative matrix factorization (NMF) \cite{lee1999learning} model to capture deeper harmonic structures in sources. Both AuxIVA and ILRMA adopt the rank-1 spatial model to enable efficient separation, where each source’s spatial image is modeled as a scaled steering vector. However, these algorithms will remain effective only when the Short-Time Fourier Transform(STFT) window fully encompasses the dominant part of the acoustic impulse response(AIR). Once this condition is violated, e.g., in highly reverberant environments, their performance quickly degrades.

Recently, convolutive transfer function-based multichannel non-negative matrix factorization (CTF-MNMF)~\cite{wang2022convolutive, wang2024semi, 10447429} has shown superior separation performance effectively, especially in highly reverberant conditions. By explicitly modeling multi-frame correlations of delayed source signals, the convolutive transfer function(CTF) model retains a finite set of delayed taps of the CTF filter in the STFT domain. Therefore, early reflections are integrated into an extended instantaneous mixing matrix, which enables efficient instantaneous BSS updates while modeling the actual mixing process more accurately. Besides, since the CTF model can efficiently model long AIR using short-time frames, CTF-MNMF also relaxes the restriction on STFT window lengths.
However, CTF-MNMF suffers from significant computational complexity due to the introduction of additional parameters, especially since its iterative projection (IP) \cite{ono2011stable} based demixing filter update requires matrix inversion for each source. This computational burden increases substantially with longer CTF filters.

To address these computational challenges, we propose an efficient variant of CTF-MNMF \cite{wang2022convolutive} by integrating the iterative source steering (ISS) \cite{scheibler2020fast} algorithm, termed CTF-MNMF-ISS. The ISS update rule completely avoids matrix inversion, significantly reducing computational complexity. Experimental results demonstrate that the proposed CTF-MNMF-ISS achieves comparable or superior separation performance relative to the original CTF-MNMF-IP method, while substantially enhancing computational efficiency.
\section{Signal Model And Problem Formulation}
Assume that $N$ sources are recorded by $M$ microphones. For the overdetermined condition where $M>N$, the signals observed at the $m$-th microphone with time index $t$ is expressed~as
\begin{align}
x_{m}(t)=\sum_{n=1}^{N} h_{m,n} * s_{n}(t),
\label{eq:(1)}
\end{align}
where $h_{m,n}$ is the time-invariant AIR from the $n$-th source to the $m$-th microphone, $x_{m}(t)$ and $s_n(t)$ are the $m$-th microphone signal and the $n$-th source signal, respectively, and $*$ represents linear convolution. The STFT of the observed signals (\ref{eq:(1)}) is derived as a sum of linear convolutions using the CTF assumption \cite{avargel2007system, talmon2009relative, wang2023spatially, wang2022convolutive}
\begin{align}
x_{m,i,j} = \sum_{n=1}^{N} \sum_{l=0}^{L_n-1} h_{m,n,i,l} \, s_{n,i,j-l}
\label{eq:cmntf},
\end{align}
where $i = 1 \ldots I$ and $j = 1 \ldots J$ are the frequency index and time-frame indexes, respectively, with $I$ and $J$ being the total number of frequency bins and time frames, $x_{m, i, j}$ and $s_{n, i,j}$ are the STFTs of $x_m(j)$ and $s_n(j)$, respectively, $h_{m, n, i, l}$ is the band-to-band filter coefficient and $L_n$ is the length of the CTF filter. 
For simplicity, we rewrite (\ref{eq:cmntf}) in vector form as
\begin{align}
\mathbf{x}_{i,j} = \mathbf{\tilde{{H}}_i} \, \mathbf{s}_{i,j}, 
\label{eq:xft}
\end{align}
where 
\begin{align*}
\mathbf{x}_{i,j} 
&= [x_{1,i,j},\, x_{2,i,j},\, \cdots,\, x_{M,i,j}]^{\top} \in \mathbb{C}^{M \times 1}, \\[4pt]
\mathbf{h}_{n,i,l} 
&= [h_{1,n,i,l},\, h_{2,n,i,l},\, \cdots,\, h_{M,n,i,l}]^{\top} \in \mathbb{C}^{M \times 1}, \\[4pt]
\mathbf{H}_{n,i} 
&= [\mathbf{h}_{n,i,0},\, \mathbf{h}_{n,i,1},\, \cdots,\, \mathbf{h}_{n,i,L-1}] \in \mathbb{C}^{M \times L_n}, \\[4pt]
\mathbf{\tilde{{H}}_i}
&= [\mathbf{H}_{1,i},\, \mathbf{H}_{2,i},\, \cdots,\, \mathbf{H}_{N,i}] \in \mathbb{C}^{M \times L}, \\[4pt]
\tilde{\mathbf{s}}_{n,i,j} 
&= [s_{n,i,j},\, s_{n,i,j-1},\, \cdots,\, s_{n,i,j-L_n+1}] \in \mathbb{C}^{1 \times L_n}, \\[4pt]
\mathbf{s}_{i,j} 
&= [\tilde{\mathbf{s}}_{1,i,j},\, \tilde{\mathbf{s}}_{2,i,j},\, \cdots,\, \tilde{\mathbf{s}}_{N,i,j}]^{\top} \in \mathbb{C}^{L\times 1}.
\end{align*}
Here, $\mathbf{H}_{i}$ is the mixing matrix for the $i$-th frequency bin, 
$\mathbf{s}_{i,j}$ stacks the delayed source signals, and $(\cdot)^{\top}$ denotes the transpose. 
Following \cite{wang2022convolutive}, we set $L=\sum_{n=1}^{N} L_{n}=M$. 
This choice makes $\mathbf{H}_{i}\in\mathbb{C}^{M\times M}$ square and full-rank, a prerequisite for the IP update rule, which requires matrix inversion at every frequency bin. 
Consequently, the demixing matrix can be defined as $\mathbf{W}_{i}=\mathbf{H}_{i}^{-1}$ as
\begin{align*}
\mathbf{W}_i
&=\bigl[\,\tilde{\mathbf{W}}_{1,i},\,\cdots,\,\tilde{\mathbf{W}}_{N,i}\bigr]^{\!H}
  \in \mathbb{C}^{L \times M}
\end{align*}
where 
\begin{align*}
\tilde{\mathbf{W}}_{n,i}
  = \bigl[\,\mathbf{w}_{n,0,i},\,\cdots,\,\mathbf{w}_{n,L_n-1,i}\bigr]^{\!H}
  \in \mathbb{C}^{M \times L_n}
\end{align*}
is the group of filters corresponding to source $n$ and each \(\mathbf{w}_{n,l,i}\) is an \(M\)-dimensional column vector. $(\cdot)^{H}$ stands for Hermitian transpose.
Now, the demixing process is denoted as
\begin{align}
y_{n,i,j,l} =  \mathbf{w}_{n,l,i}^{{H}} \mathbf{x}_{i,j} \quad \text{for } n = 1, \dots, N,
\label{eq:demix}
\end{align}
where
$y_{n,i,j,l}$ is the estimated source signal with $l$ taps delay. Each source is modeled as a complex Gaussian random variable with zero mean and time-varying variance $\lambda_{n, i,j}$. The power spectral density (PSD) is represented using NMF as~\cite{sekiguchi2020fast}
\begin{equation*}
\lambda_{n,i,j-l} = \sum_{k=1}^{K_n} b_{n,i,k}\,v_{n,k,j-l},
\end{equation*}
where $b_{n,i,k}$ and $v_{n,k,j-l}$ are the NMF basis and activation components for the $n$-th source, respectively, with $k=1,\dots,K_n$, where $K_n$ is the number of latent spectral bases for source $n$.

The objective function is obtained by calculating the negative log-likelihood function as in \cite{wang2022convolutive}:
{\small
\begin{align}
\mathcal{L}(\mathbf{W},\boldsymbol{\Lambda}) = \sum_{i,j,n,l}  \left( \log \lambda_{n,i,j-l} + \frac{|y_{n,i,j,l}|^2}{\lambda_{n,i,j-l}} \right) \nonumber\\- 2J \sum_{i} \log |\det \mathbf{W}_i| + cst. \label{eq:loss}
\end{align}
}
The task is now transformed into minimizing the objective function (\ref{eq:loss}) with respect to the demixing matrices $\{\mathbf{W}_i\}$ and PSD parameters $\{\lambda_{n,i,j}\}$, which yields the following optimization formulation:
\begin{align}
\{\mathbf{W}^\star,\boldsymbol{\Lambda}^\star\}
  &= \arg\min_{\mathbf{W},\boldsymbol{\Lambda}}
     \,\mathcal{L}(\mathbf{W},\boldsymbol{\Lambda}).
\end{align}
\section{Proposed Method}
\subsection{Optimization algorithm}
\subsubsection{Update of  $\mathbf{W}_i$}
In the following, we deduce update rules for optimizing~(\ref{eq:loss}) based on the auxiliary function technique~\cite{ono2011stable}. It can be obtained that
{\small
\begin{align}
\mathcal{L}^{+}
  = -2 \sum_{i=1}^{I} \log |\det \mathbf{W}_{i}|
    + \sum_{i=1}^{I} \sum_{n,l}
      \mathbf{w}_{n,l,i}^{H}\,\mathbf{Q}_{n,l,i}\,\mathbf{w}_{n,l,i},
\label{eq:aux}
\end{align}
}
where the weighted covariance matrix $\mathbf{Q}_{n,l,i}$ is defined as
\begin{align}
\mathbf{Q}_{n,l,i}
  = \frac{1}{J}\sum_{j=1}^{J}
    \frac{\mathbf{x}_{i,j}\,\mathbf{x}_{i,j}^{H}}{\lambda_{n,i,j-l}}.
\end{align}
The original IP-based Optimization goes as in \cite{wang2022convolutive}
{\small
\begin{align}
\mathbf{w}_{n,l,i}
 &\leftarrow \bigl(\mathbf{W}_i\,\mathbf{Q}_{n,l,i}\bigr)^{-1}\mathbf{e}_{\,(L_{1}+L_{2}\cdots+L_{n-1}+l+1)},\label{eq:ipinvserse}\\[4pt]
\mathbf{w}_{n,l,i}
 &\leftarrow \mathbf{w}_{n,l,i}
    \bigl(\mathbf{w}_{n,l,i}^{H}\mathbf{Q}_{n,l,i}\mathbf{w}_{n,l,i}\bigr)^{-1/2},
\end{align}
}
where $\mathbf{e}_{\,(L_{1}+L_{2}\cdots+L_{n-1}+l+1)}$ is a unit column vector whose $(L_{1}+L_{2}\cdots+L_{n-1}+l+1)$th element equals to one. To simplify the ISS update formulation, we flatten the two-dimensional index $(n,l)$ into a single index $r = (L_{1}+L_{2}\cdots+L_{n-1})+l+1$, where $n=1,\ldots,N$, $l=0,\ldots,L_n-1$, and $r = 1,\ldots, L$. Hence, $\mathbf{w}_{n,l,i}$ can be denoted as $\mathbf{w}_{r,i}$ for notational convenience. 
The proposed method employs the ISS updating rule \cite{scheibler2020fast}, which updates the entire filter using a rank-1 matrix as
{\small
\begin{align}
\mathbf{W}_i \leftarrow \mathbf{W}_i - \mathbf{z}_{r,i}\,\mathbf{w}_{r,i}^{H},
\label{eq:update}
\end{align}}
where $r$ is the index indicating the rank-1 updates applied sequentially to each source and delay tap, and $\mathbf{z}_{r,i} = [z_{1,r,i},\, \dots,\, z_{L,r,i}]^{\top} \in \mathbb{C}^{L \times 1}$ is a vector to estimate and $(\cdot)^{*}$ denotes complex conjugate. 

To derive the optimal update direction, we substituting the rank-1 update~(\ref{eq:update}) into the auxiliary objective function~(\ref{eq:aux}), we get the new optimization objective:
{\small
\begin{align}
\mathcal{L}_{\mathrm{ISS}}(\mathbf{z}_{r,i}) 
&= -2 \sum_{i=1}^{I} \log \left| \det \left( \mathbf{W}_i - \mathbf{z}_{r,i}\,\mathbf{w}_{r,i}^{H} \right) \right|+ \sum_{i}\sum_{p=1}^L  \nonumber \\
& 
\left( \mathbf{w}_{p,i} - z_{p,r,i}^{*}\,\mathbf{w}_{r,i} \right)^{H}
\mathbf{Q}_{p,i}
\left( \mathbf{w}_{p,i} - z_{p,r,i}^{*}\,\mathbf{w}_{r,i} \right),
\end{align}
}
where $p$ is a dummy index ranging from 1 to $L$ that enumerates all rows, by the matrix determinant lemma $\det\bigl(\mathbf A - \mathbf u\mathbf z^{ H}\bigr)= \det(\mathbf A)\bigl(1 - \mathbf z^{H} \mathbf A^{-1} \mathbf u\bigr)$,
we have
\begin{align}
\det\left( \mathbf{W}_i - \mathbf{z}_{r,i} \mathbf{w}_{r,i}^{H} \right)
&= \det\left( \mathbf{W}_i \right) \left( 1 - z_{r,r,i} \right).
\end{align}
Taking the derivative of $\mathcal{L}$ with respect to $z_{p,r,i}^{*}$ and setting it to zero, we consider two cases.\\
First, when $r \ne p$, we can obtain
\begin{equation}
\frac{\partial \mathcal{L}_{\mathrm{ISS}}}{\partial z_{p,r,i}^{*}} =
- \mathbf{w}_{r,i}^{H} \mathbf{Q}_{p,i} \mathbf{w}_{p,i}
+ z_{p,r,i} \, \mathbf{w}_{r,i}^{H} \mathbf{Q}_{p,i} \mathbf{w}_{r,i}
\end{equation}
Next, when $r = p$, the partial derivative of $\mathcal{L}_{\mathrm{}}$ is
\begin{equation}
\frac{\partial \mathcal{L}_{\mathrm{ISS}}}{\partial z_{p,r,i}^{*}} =
\frac{2}{1 - z_{r,r,i}} -
2(1 - z_{r,r,i})\, \mathbf{w}_{r,i}^{{H}} \mathbf{Q}_{r,i} \mathbf{w}_{r,i}
\end{equation}
Setting these expressions to zero yields the closed-form solution, then the update of $z_{r,p,j}$ can be obtained that
\begin{align}
z_{p,r, i}
=\begin{cases}
\dfrac{\mathbf w_{r,i}^{ H}\mathbf Q_{p,i}\mathbf w_{p,i}}
      {\mathbf w_{r,i}^{ H}\mathbf Q_{p,i}\mathbf w_{r,i}},
 & p\neq r,\\[10pt]
1-\bigl(\mathbf w_{r,i}^{ H}\mathbf Q_{r,i}\mathbf w_{r,i}\bigr)^{-1/2},
 & p = r.
\end{cases}
\label{eq:v_closed}
\end{align}
Then, by using the original ISS update rule, the demixing matrix $W$ is solved.
\subsubsection{Update of $\boldsymbol{\Lambda}$}
With $\mathbf W$ fixed, minimising $\mathcal L$ over
$\{b_{n, i,k},\,v_{n,k,j}\}$ is equivalent to minimising a sum of Itakura–Saito divergences between the $|y_{n,i,j,l}|^{2}$ and $\lambda_{n,i,j-l}$. Using a Majorize-Minimization (MM) framework yields the following Multiplicative Update (MU) rules \cite{wang2022convolutive}.
For each $n, i,k$, the update of the basis and the activation can be obtained by 
\begin{align}
b_{n,i,k} &\leftarrow b_{n,i,k} 
\sqrt{
\frac{
\sum_{j=1}^{J} \sum_{l=0}^{L-1} \left| y_{n,i,j,l} \right|^2 v_{n,k,j-l} \lambda_{n,i,j-l}^{-2}
}{
\sum_{j=1}^{J} \sum_{l=0}^{L-1} v_{n,k,j-l} \lambda_{n,i,j-l}^{-1}
}
},
 \label{eq:B_update}\\
v_{n,k,j} &\leftarrow v_{n,k,j} 
\sqrt{
\frac{
\sum_{i=1}^{I} \left| y_{n,i,j,0} \right|^2 b_{n,i,k} \lambda_{n,i,j}^{-2}
}{
\sum_{i=1}^{I} b_{n,i,k} \lambda_{n,i,j}^{-1}
}
}.
 \label{eq:V_update}
\end{align}
To resolve scale ambiguity among $b_{n,i,k}$, $v_{n,k,j}$ and the demixing matrix $\mathbf{W}$, we compute the average power:
\begin{align}
  \mu_{n,l} = \sqrt{\frac{1}{I J}\sum_{i,j} |y_{n,i,j,l}|^{2}},
\end{align}
then apply rescaling:
\begin{align}
y_{n,i,j,l} &\leftarrow y_{n,i,j,l} \, \mu_{n,l}^{-1}, \\
b_{n,i,k} &\leftarrow b_{n,i,k} \, \mu_{n,0}^{-2}.
\end{align}
The same scaling is also applied to the corresponding rows of $\mathbf{W}$ to ensure consistent signal energy.
\subsection{Source image estimation}
To avoid the spatial distortion caused by using~(\ref{eq:demix}) when estimating the source signal in a reverberant environment, after estimating the demixing matrix and the source PSDs, we reconstruct the spatial images $\hat{\mathbf{c}}_{n,i,j} \in \mathbb{C}^{M}$ using a multichannel Wiener filter (MWF) \cite{sawada2013multichannel}. The goal is to minimize the mean squared error (MSE) between the estimated and true spatial images:
\begin{align}
\mathbf{M}_{n,i,j}^{\mathrm{opt}} = \arg\min_{\mathbf{M}_{n,i,j}} \mathbb{E} \left[ \left\| \mathbf{c}_{n,i,j} - \mathbf{M}_{n,i,j} \mathbf{x}_{i,j} \right\|_2^2 \right],
\end{align}
where $\mathbf{c}_{n,i,j} \in \mathbb{C}^{M}$ is the source image. The optimal estimator is given by:
\begin{align}
\mathbf{M}_{n,i,j}^{\mathrm{opt}} = \mathbb{E}[\mathbf{c}_{n,i,j} \mathbf{x}_{i,j}^{{H}}] \cdot \mathbb{E}^{-1}[\mathbf{x}_{i,j} \mathbf{x}_{i,j}^{{H}}].
\label{mwf}
\end{align}
Under the proposed CTF-based spatial model, (\ref{mwf}) becomes
\begin{align}
\hat{\mathbf{c}}_{n,i,j} &= \mathbf{M}_{n,i,j}^{\mathrm{opt}} \mathbf{x}_{i,j}
\nonumber\\&= \mathbf{H}_{n,i} \boldsymbol{\Lambda}_{n,i,j} \mathbf{H}_{n,i}^{{H}} 
\left( \mathbf{H}_i \boldsymbol{\Lambda}_{i,j} \mathbf{H}_i^{{H}} \right)^{-1} \mathbf{x}_{i,j},
\end{align}
where $\boldsymbol{\Lambda}_{n,i,j}\in\mathbb{R}^{L \times L}$ is the diagonal PSD matrix of source $n$. Therefore, the MWF serves as the final stage to recover the spatial images.

\section{Computational Complexity Analysis}\label{sec:complexity}
This section compares the computational costs of CTF-MNMF-IP and CTF-MNMF-ISS with respect to the demixing matrix $\mathbf{W}{i}$. As both methods employ identical MU rules for the NMF parameters, the computational complexity of this part is the same. Thus, the difference in the overall computational burden stems solely from the optimization strategy applied to $\mathbf{W}{i}$.

CTF-MNMF-IP updates the demixing matrix $\mathbf{W}_{i}$ by solving a full linear system. Inverting an $L\times M$ matrix  in~(\ref{eq:ipinvserse}) requires $\mathcal{O}(M^{3})$ floating-point
operations. Repeating this for all signals across $I$ frequency bins and $L$ sources yields
\begin{align}
\mathcal{C}_{\mathrm{IP}}
  &\propto
  \mathcal{O}\!\bigl(ILM^{3}\bigr). \label{eq:cip}
\end{align}

CTF-MNMF-ISS rewrites the update as a rank-one steering step so the costly inversion disappears. The dominant operation becomes the matrix–vector product plus rank-one correction, whose computational complexity scales quadratically as $\mathcal{O}(M^{2})$.
Accounting again for all signals and $I$ frequencies gives
\begin{align}
\mathcal{C}_{\mathrm{ISS}}
  &\propto
  \mathcal{O}\!\bigl(ILM^{2}\bigr). \label{eq:ciss}
\end{align}
which is an order of magnitude lower than that of the IP method.
\section{Experiment}
In this section, we will compare the performance of CTF-ISS with the traditional method in \cite{wang2022convolutive}. For simplicity, we omit MNMF in this section. 
\subsection{Experimental setup}
The observation signals are generated by convolving speech signals from the TIMIT database \cite{garofolo1993timit}. Each mixed signal consists of two speech segments, randomly selected from different speakers, concatenated to form an 8-second clean speech signal. To validate the performance under realistic reverberant environments, we utilize impulse responses from the RWCP dataset, specifically E2A with a reverberation time of $\mathrm{RT}_{60}=300$ ms, JR2 with $\mathrm{RT}_{60}=470$ ms, and E2B with $\mathrm{RT}_{60}=1300$ ms.

\begin{figure} 
\centering 
\includegraphics[width=0.45\textwidth]{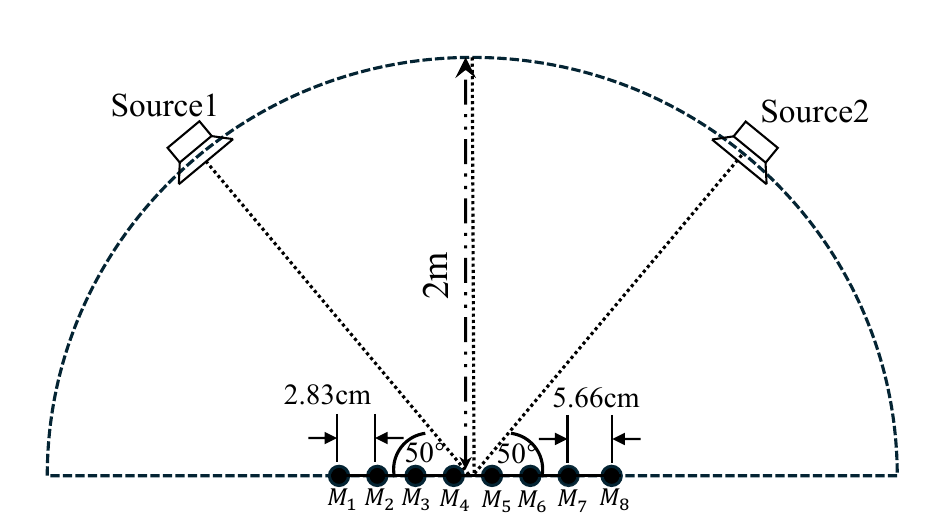} 
\caption{Illustration of the simulation setup.} 
\label{experiment setting} 
\end{figure}

The geometric configuration used in our experiments is illustrated in Fig.~\ref{experiment setting}. Two sound sources are positioned 2 meters away from the microphone array center, forming an angular separation of $100^\circ$ (at azimuth angles $-50^\circ$ and $+50^\circ$). The microphone array consists of eight omnidirectional microphones ($M_1$ to $M_8$) arranged in a linear formation. The experimental settings are identical to those of the conventional method. The specific microphone configurations for different array sizes utilized in this study is summarized in Table I.
All recordings are sampled at 16 kHz. The time-frequency representation is obtained using the STFT with a 1024-point Hann window and a hop size of 25\%. All of the separation matrices $\mathbf{W}_i$ are initialized as identity matrices, and the number of iterations is fixed at 100. The number of bases $K_n$ is set to 3. The number of microphones used varies among 4, 6, and 8. Each microphone is regarded as an independent observation channel, and two sources are active in all scenarios. The CTF filter length $L_n$ is set to 2, 3, and 4 taps for the 4-, 6-, and 8-channel microphone arrays, respectively, as shown in Table~\ref{tab:mic_config}. All experiments are conducted on a laptop with an AMD Ryzen 7 5800H CPU.

\begin{table}[t]
\centering
\caption{Microphone configurations and filter lengths}
\label{tab:mic_config}
\begin{tabular}{c c l}
\toprule
$M$ & Filter length & Microphones configurations \\
\midrule
4 & $L_n = 2$ & $M_3,\, M_4,\, M_5,\, M_6$ \\
6 & $L_n = 3$ & $M_2,\, M_3,\, M_4,\, M_5,\, M_6,\, M_7$ \\
8 & $L_n = 4$ & $M_1,\, M_2,\, M_3,\, M_4,\, M_5,\, M_6,\, M_7,\, M_8$ \\
\bottomrule
\end{tabular}
\end{table}
\subsection{Experimental results}
\begin{figure}[t] 
\centering 
\includegraphics[width=0.4\textwidth]{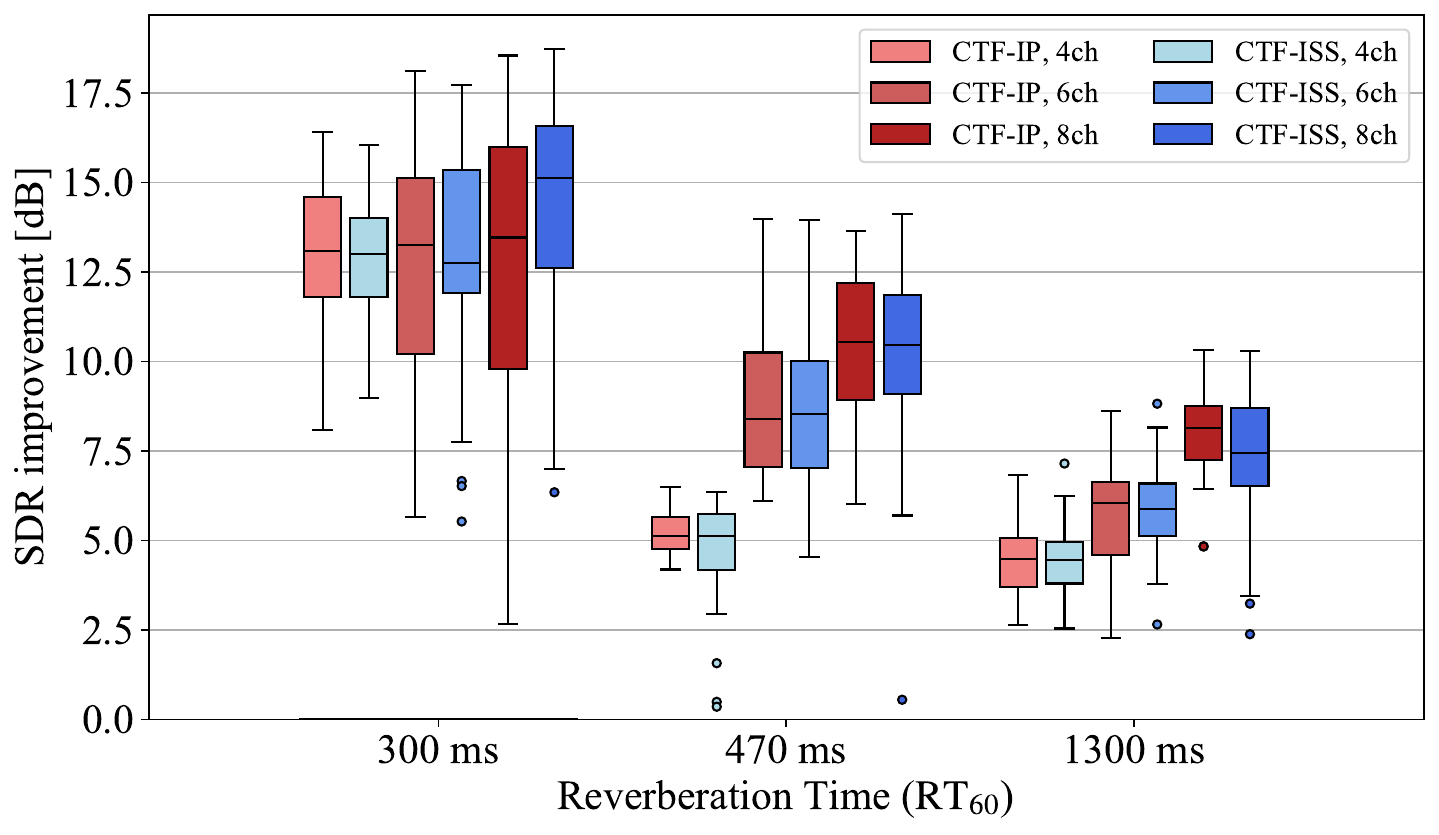} 
\caption{Average SDR improvement under different reverberation and microphone setups.} 
\label{experiment result300} 
\end{figure}
Figure~\ref{experiment result300} compares CTF-IP (red) and the proposed CTF-ISS (blue) in terms of SDR-improvement for three reverberation times ($\mathrm{RT}_{60}=300$, $470$, and $1300$ ms). Under all conditions, the median SDR decreases with increasing reverberation time, demonstrating the inherent difficulty of source separation in highly reverberant environments. Increasing the number of microphones consistently improves SDR performance by around 2–4 dB, with the most notable improvement observed at 1300 ms, where an 8-channel configuration outperforms the 4-channel case by nearly 4 dB.

Regarding the optimization strategy, CTF-ISS maintains comparable median performance to CTF-IP, with differences never exceeding 1 dB. However, CTF-ISS exhibits narrower inter-quartile ranges and fewer outliers, particularly noticeable at longer reverberation times of 470 ms and 1300 ms. This improvement indicates that ISS updates provide robustness and lessen sensitivity to initial conditions, yielding more consistent separation performance.
\begin{figure} 
\centering 
\includegraphics[width=0.4\textwidth]{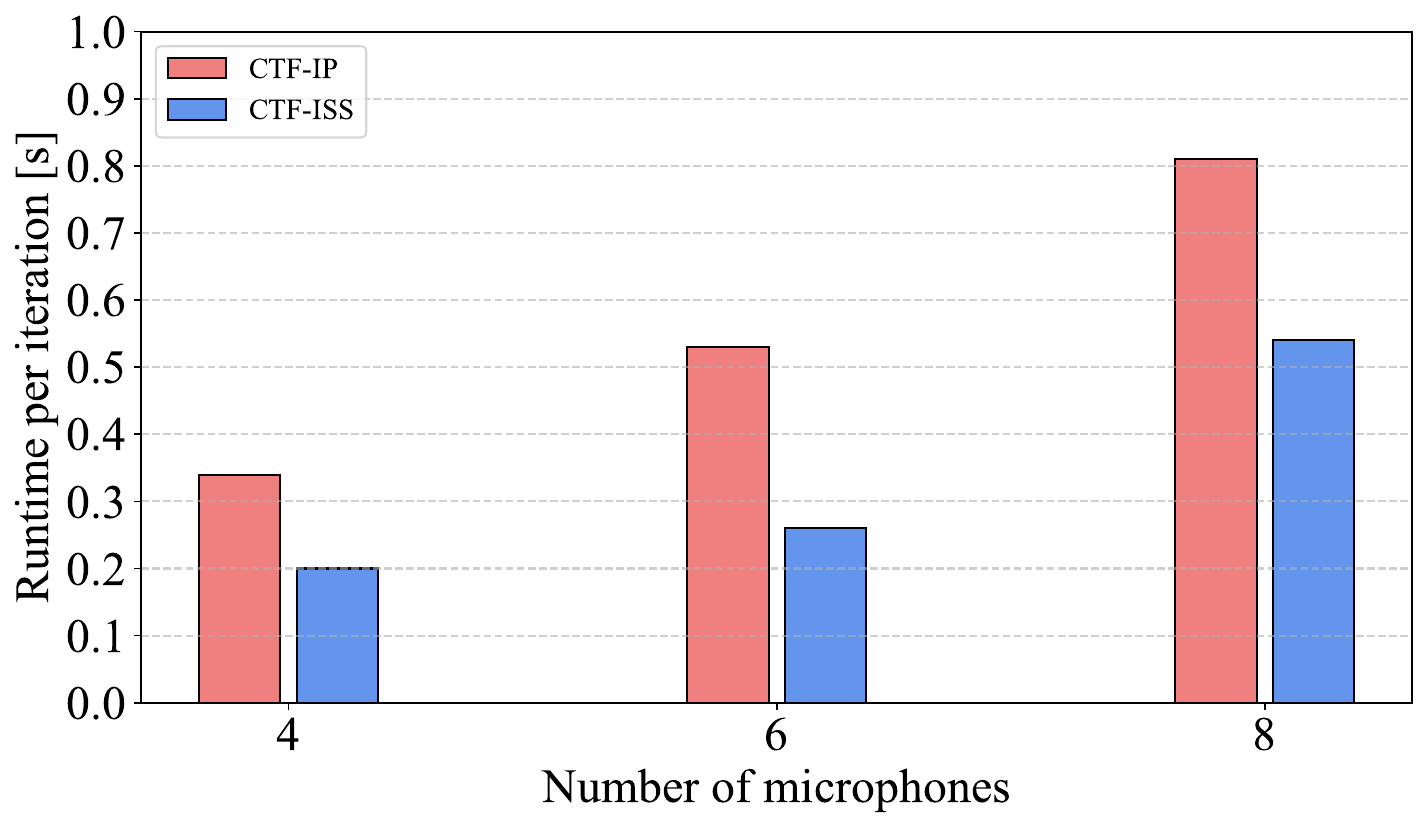} 
\caption{Average runtime under different microphone setups.} 
\label{spatial_runtime} 
\end{figure}
Fig~\ref{spatial_runtime} further analyzes the computational efficiency of the two methods by presenting their average runtime across different microphone configurations. The proposed CTF-ISS consistently demonstrates lower computational costs compared to CTF-IP. Specifically, CTF-ISS achieves runtime reductions of approximately 41\% , 51\%, and 33\%, respectively, for the 4-channel, 6-channel, and 8-channel setups. These significant reductions underscore the computational advantage of the ISS-based approach, which is particularly beneficial for real-time processing applications or systems with limited computational resources.
Besides the CPU saving, ISS avoids repeated matrix inversions,
leading to smaller memory usage and better numerical stability.
Consequently, the proposed CTF–ISS variant achieves essentially the same separation accuracy as CTF–IP while running significantly faster and exhibiting lower run-to-run variance.
The experimental results confirm that the proposed CTF-MNMF-ISS method not only preserves high-quality source separation comparable to conventional method but also offers enhanced stability and significantly reduced computational complexity, making it particularly advantageous as the number of microphones increases or under challenging reverberant conditions.
\section{Conclusions}
In this paper, we proposed an accelerated CTF-MNMF algorithm for overdetermined blind source separation, named CTF-MNMF-ISS. By integrating the iterative source steering approach, we successfully circumvented the numerical instability associated with iterative projection updates, thus significantly enhancing the practicality and robustness of the original algorithm.
CTF-MNMF-ISS substantially reduces computational complexity, achieving a runtime reduction of approximately 40\%. It simultaneously maintains or even surpasses the separation performance of the original IP-based method. Furthermore, the ISS-based updates exhibited greater numerical stability and robustness against variations in initialization, particularly beneficial in highly reverberant environments.
\pagestyle{fancy}
\printbibliography
\end{document}